**Title**

"iNaturalist citizen science community during City Nature Challenge: new computational approach for analysis of user activity"

**Authors:**

Liubov Tupikina (Université de Paris, INSERM U1284, Center for Research and Interdisciplinarity (CRI); 0000-0002-7169-5706, CorrelAid), Frank Schlosser (Institute for Theoretical Biology, Humboldt University of Berlin, D-10115 Berlin, Germany; orcid 0000-0003-1649-4300), Vadim Voskresenskii (Free University of Berlin, 0000-0001-7284-8527), katharina kloppenborg (Université de Paris, INSERM U1284, Center for Research and Interdisciplinarity (CRI); orcid 0000-0002-5364-8965), Florence Lopez (scieneers GmbH orcid 0000-0002-7331-2289), Albrecht Mariz (CorrelAid; orcid 0000-0003-1558-2261), Anna Mogilevskaja (CorrelAid), Muki Haklay (UCL, 0000-0001-6117-3026), Bastian Greshake Tzovaras (Université de Paris, INSERM U1284, Center for Research and Interdisciplinarity (CRI), Open Humans Foundation; orcid 0000-0002-9925-9623)

**Abstract**

**Keywords:** citizen science community, networks analysis, clustering, spatial-temporal patterns

Analysing patterns of engagement among citizen science participants can provide important insights into the organisation and practice of individual citizen science projects. In particular, methods from statistics and network science can be used to understand different types of user behaviour and user interactions to help the further implementation and organization of community efforts. Using publicly available data from the iNaturalist community and their yearly City Nature Challenges (CNC) from 2017-2020 as an example; we showcase computational methods to explore the spatio-temporal evolution of this citizen science community that typically interacts in a hybrid offline-online way. In particular, we investigate the user types present in the community along with their interactions, finding significant differences in usage-behavior on both the level of engagement and the types of community tasks/roles and how they interact with the network of contributors. We expect that these computational analysis strategies will be useful to gain further understanding of other citizen science communities and projects.

**Introduction**

Crowdsourcing – "an online, distributed problem-solving and production model that leverages the collective intelligence of online communities to serve organizational goals" (Brabham 2013) – is entering its second decade. Many crowdsourcing efforts are applied to problems where the data includes a significant geographical element - that is, the geographical information is necessary in the process of solving the specific problem.

From Uber to dockless bikes, food delivery, or updating Google Maps, the procedures and techniques to its use are now commonplace.

Beyond commercial crowdsourcing, there are also open, non-commercial efforts that are well aligned with open science principles (Peters 2014). These systems are aimed at creating an open accessible body of high quality knowledge that can be used by anyone, ranging from commercial purposes to humanitarian applications. One of the most well-known examples of such open geographical crowdsourcing efforts is the OpenStreetMap (OSM) project, in which participants create an open digital map of the world (Haklay and Weber 2008). At the same time, OSM and other similar efforts can be viewed as a form of citizen science, which we define here as the participation of non-professional people in knowledge production. Various citizen science projects around the world actively engage individuals in scientific research providing them with information about types of participation and data collection methodology (Kloppenborg, Ball and Greshake Tzovaras 2021). However there are various concerns raised about the crowdsourcing and citizen science projects, such as questions about data quality (Anon 2015), the motivations of participants, which is related to users' attrition and knowledge production.

The iNaturalist platform is a widely used citizen science project (Wittmann, Girman and Crocker 2019). iNaturalist provides possibilities for people to upload and review their observations from nature, connects naturalists with other users who are experts in classification and who can identify the observed organisms, helps scientists and resource managers understand when and where organisms occur.

Users upload observations or identifications of species, while also being allowed to submit aggregated observations from others. When a user uploads their observations to the platform, they provide a guess on the species. Other users then participate in a communal validation process by adding their own identifications of the species, until the label of an observation is changed to a research grade once the observation receives more than two independent classifications.

Being a social network and online platform for sharing, iNaturalist allows mapping and identifying biodiversity observations around the globe. While building a scientifically valuable biodiversity database via this approach, they primarily aim to connect people to nature and encourage biodiversity protection, including organisation of various localised events. Of particular interest are City Nature Challenges (CNC) which are organized by the iNaturalist community. As the CNC are bottom-up, voluntarily organised global events that have grown significantly within five year, they are of special interest. The first CNC started in 2016 as a friendly competition between the San Francisco and Los Angeles natural history museums (Nugent 2018). It is a weekend-long effort during spring, to record as many observations of species of all kinds within a given territory. By 2017 16 cities across the US participated in the event, and by 2018 it spread to Europe, South America, and Asia with 68 cities. In 2020, 244 cities took part across the world.

Analysing the patterns of engagement in citizen science can provide an important insight into the organisation and practice of citizen science activities. The iNaturalist project is very loosely coordinated, due to resource constraints (Nugent 2018). As a result, each of the participating cities is free to organise the activities in any form that fits the goals of local organisers, their abilities, and their networks. While most cities use iNaturalist, local variations of data collection platforms are permitted. This means that by analysing patterns of contribution across different cities, we can additionally learn about different modes of bottom-up, limited budget, city-scale events.

Inspired by similar work, which has looked at patterns of contributions to OSM (Schmidt and Klettner 2013) and eBird (Rutter et al. 2021), we showcase how data science methods can be applied to citizen science platforms to explore patterns of contributions made by participants of the iNaturalist citizen science platform during the CNCs in Los Angeles, San Francisco, and London. In particular, we explore how methods from statistics and network science can be used to understand different types of user behaviour and user interactions to provide strategies that can be applied to other online citizen science efforts.

**Data overview**

We downloaded data from iNaturalist using the platform's API, focusing on data from three cities and the time periods during which CNCs happened: London (2018-2020), Los Angeles (2017-2020) and San Francisco (2017-2020). Part of the extracted and used data is available on GitHub.

This data from iNaturalist includes information about users, their observations and interactions between users through comments on observations they made. In particular, the data contain the information about users ids, time and location, when a user submitted the observation, the identified species and whether other users agree or disagree with the identification, their first guess of the observed species, as well as identification corrections. The reference to the API description from iNaturalist platform describes the data obtained directly from the platform.

**General statistics**

Our data set includes a total of 244,484 observations that were uploaded to iNaturalist platform during the 11 CNCs in 3 different cities over 3 years, by 11,300 distinct users, 2,142 of which participated in more than one challenge.

The distribution of contributions is highly skewed to the left, **(Figure 1 a),** with 34.6% of the participants uploading one single observation per challenge, and 62.7% uploading one to five observations. The top 50% of participants contribute 97% of the observations, the top 10% - 77% of the observations and the top 1% - 41% of the observations (**Figure 1 b**). The highest number of observations a single user contributed to one CNC is 1,414 observations in Los Angeles 2018, the highest number for one user summed up over several challenges is 3,572.

In **Figure 2,** we show user activity per city/year. Overall we see that across all cities, the number of observations was fluctuating but roughly stayed the same **(Figure 2 a)**. In contrast, the number of users increased over time **(Figure 2 b)**, such that consequently the average number of observations per user decreased over the years.

**Computational methods and results**

We adapt and apply computational methods to the dataset from iNaturalist. We start first with classification methods of users, as one of the most general characteristics of activity patterns. From that we move to application of general statistical methods to characterize users' attrition. Then we extend our analysis of users' activity and use geospatial analysis to process and visualise the geographical information about citizen science data. Finally, we also apply network projections to our dataset to get the understanding of the citizen science community from a social network view point.

*Users' classification*

Users can show different kinds of behavior in how they interact during the CNC. We used the information on *identifications* and *observations* to classify users' behavior in four broad groups or clusters: *observers*, who mostly submitted observations; *identifiers*, who mostly helped to identify the behavior of other users; *high activity* users, who submitted a high number of observations and identifications, and *low activity* users, see **Figure 3**. We find that, across all cities and years according to the classification we propose, most users are low activity users (58%, 13,247). Around 37% of users are either observers (25%, 5,637) or identifiers (12%, 2,821), while only a small fraction are high activity users (5%, 1,122). We also observe that only most people are low activity users, but even among those who are, the majority tends to provide more data, while a much smaller amount is providing 'maintenance' or identifications.

For the classification we applied the k-means clustering algorithm in the two-dimensional space of observations and identifications (Hartigan and Wong 1979). We determined the optimal number of four clusters using the elbow method. We applied the clustering to each CNC separately, so that the same user can exhibit a different type of behavior over the years. Furthermore we use this information for social network analysis to study behaviour of users belonging to these various groups.

*Users' attrition*

An important question for the organization of citizen science events in general, and for iNaturalist challenges in particular is, how new users can be onboarded most successfully. Overall, in our data set, there are 11,300 distinct users, of which 2,142 participated in more than one challenge. In order to better understand the users' attrition phenomenon (when users are leaving the platform), we tried to answer the main question: How different are the attrition (or retention) dynamics of those users who have joined the platform via a challenge and regular platform users or those who contribute to the platform in periods different from challenge periods?

We analyze how long new users stay on the platform after joining it for 6 months. The comparison of attrition patterns of 'challengers' (users participating in a challenge) and non-challenge users (users who joined the platform in a different time than during the challenge) was made only for San Francisco in 2019 and 2020. To identify both groups of users, we collected observations for San Francisco from the beginning of 2017 until 2020 for both: users who joined the platform during the challenge and during the neighboring time period. We define new regular users as those who have joined the platform any day of 2019 or 2020 that is not related to the challenge; in turn, new users of the challenge are users who joined the platform during the challenge. Some first observations are:

- A high number of users in 2019 are new, since they did not participate in 2018.
- Around 15-20% of all new users in 2019 participate in the next year, 2020.

In order to explain the turn-over of users we looked into how users structured their additional overall trends. These trends explain why studying users' attrition is important for citizen science, as well as for other communities. We find that users make fewer observations in the first year, in their second year, successfully onboarded users make more observations than average users even though observations decline in general.

In **Figure 4** we show the dynamics of the attrition for two types of users: (left part of both figures) for participants of City Nature Challenge and (right part of both figures) for regular users. In both charts the X-axis shows the number of months users were active after joining, and the Y-axis stands for the percentage of users. Based on the charts, we can see a tendency that users who joined the platform outside of CNC stay for longer periods on the platform than the participants who joined during CNC. This difference could indicate different intrinsic motivations of these two groups of users: regular users might be more interested in the affordances of the platform, which could make them stay for a longer period, while challenge users, supposedly, might rather join the platform for the specific challenge event.

It's also interesting that the regular users, who joined in 2020 (see **Figure 4**, bottom), stayed for longer periods on the platform than regular users in 2019 (see **Figure 4,** top).

Testing potential reasons for this would need more information about users' intrinsic motivation and probably also additional survey data regarding this.

### *Geospatial analyses*

For deeper understanding of the temporal evolution of community in iNaturalist, we investigate potentially related factors, such as properties of spatial distribution of citizen science communities. It is possible to do so with accessing additional geospatial data about each city, such as greenspaces, blue spaces (lakes, rivers etc.) and information about the population of an area. For this we used data provided from Geofabrik [Geofabrik] and an API [OSMnx] to extract the data from OSM sources [Github correlaid].

One of the questions we asked was whether the surroundings of a user and the properties of their environment affects the amount and quality of contributions they are making. Additionally, we explored whether different areas in a city have an impact on the species that are observed in this city.

Overall, we focused on several different spatial aspects, the first one being where users make the observations. First observation shows that users mainly make contributions around each city, in which they participate in the challenge, and do not make observations outside of the city area, e.g. in the city suburban areas. We also looked at the travelling patterns of users, meaning their trajectories. We assumed that the amount of travelling a user does would give us useful information about the amount of contribution they make and help us to establish a classification of users into different spatial groups, e.g. groups of people who make observations in green or blue spaces. By combining the data from iNaturalist with data from OSM we analyze the spatial patterns of the observations in relation to greenspaces and blue spaces.

From our calculations we see that in 2019 (pre-COVID) there were substantially more observations in the centre of the city of London, while in 2020 the observations started to be more spread-out around the city itself. This can be observed by looking at the quadrant count, as depicted in **Figure 5**.

In **Figure 6** we depicted London Hyde Park as an example of the shift of distributions from greenspaces to non-green spaces. In 2019, the park was an observational hot-spot, meaning that many people were doing their observations there. Whereas in 2020, probably due to COVID lock-down effect, only a very small amount of the observations made in London, came from Hyde Park. This phenomenon is a very good example for the shift that took place between observations made in 2019 and 2020. It can be further observed in **Figure 7** where the relative amount of observations made in greenspaces and non-green spaces is shown. We observed that during the years 2018 and 2019, the amount of observations made in greenspaces and non-green spaces was nearly equally distributed. But in 2020, a shift took place and nearly 75% of all observations made in London were located in non-green spaces. The reasons for such shifts could be verified with additional survey data and causal analysis techniques.

Additionally, we looked into the distribution of species observations across all cities and all years of challenges (2018-2020). While in the first year of the CNC the observations of all species are concentrated among certain areas (mostly around greenspaces), the geographical distribution of species observations increases in the following observation years. However, as some species, e.g. aquatic organisms can only be observed in appropriate habitats, this trend is not observed for all species and should be checked more.

## Social network analysis

The analysis of collaboration networks in citizen science projects can help to understand the collaboration processes between participants and their respective contributions. In this section we analyze the social interaction network on the iNaturalist platform that is created by user collaborations. In the network, the nodes correspond to users of the platform, and links correspond to social interactions between users. We count one social interaction between two users A and B if user A submits an observation and to which user B adds an identification, or vice versa, see **Figure 8a**. The weight of the link is given by the number of such interactions, where we do not distinguish the direction of the interaction, so that the resulting network is undirected.

The social interaction network yields interesting insights into the collaboration dynamics, see **Figure 8b.** We find that there is a large variance in how connected and how central users are in the network, which is a common observation for empirical social networks (Newman 2010). High activity users with many observations and identifications feature prominently in the center of the network, while low activity users are situated more along the periphery. Identifiers feature more prominently as they have on average more social interactions than observers, while there are more observers in the network in total, compare **Figure 3**.

There are many more advanced tools and methods from network science we can use to understand the network. We demonstrate this by investigating different centrality metrics of users in the network, see **Figure 9.** The degree centrality $x_d$ of a user corresponds to the number of social interactions with other users, i.e. the sum of all the users' links in the network (Newman 2010). We find a simple observation in that the degree centrality grows with the total number of social interactions and in particular the relation between the user activity and user centrality: low activity users have the smallest degree centrality, followed by observers, identifiers and finally high activity users, see **Figure 9a**. However, a more advanced metric such as the eigenvector centrality $x_e$ can show that the picture is more nuanced. The eigenvector centrality measures how central a user is in the network, taking also into account how central the users are that they are linked to (Bovet and Makse 2021). Here, we find that observers and identifiers have a similar centrality in the network, while high activity users clearly dominate the network, see **Figure 9b.**

Another interesting question related to social network analysis is the formation of the initial core group that often plays a crucial role for establishing sustainable connections. Interestingly the core may be completely self-emerged, e.g. without any moderators, like in other citizen science projects [Aristeidou et al. 2015]. Network analysis is one way to analyse citizen science contributions and can be compared with other classical methods.

**Discussion and Outlook**

In this paper we provided some analyses on temporal and spatial aspects of users activity and user interactions related to the City Nature Challenges of the citizen science platform iNaturalist. The patterns of engagement we found can provide an important insight into user-attrition factors, the organisation and practice of citizen science activities in general.

We have focused on a quantitative, data-driven approach, which provides insights that can be used by CNC organisers in different locations to learn about the evolution of their effort over the years, and potentially adjust their activities accordingly. A useful extension of our approach could be detailed discussions with project organisers and participants to understand their actions and the observed patterns.

One big factor that reinforces contributions of the members in a community is the presence of notifications of the members about new activities and explained need to contribute (Kraut, Resnick and Kiesler 2011). In the case of iNaturalist, the announcement of new CNC each year can be understood as a type of notification. In our paper we investigated other potentially important characteristics of users' behaviour traces, such as centrality measures of the social network constructed from the user's interactions on the platform. We tried various data-driven techniques to characterise the evolution of the iNaturalist community across various cities.

An overall trend – across all cities and years for CNC iNaturalist challenges – is that the number of observations per user decreased from 2018 to 2020. While we observe an increase in total users during the same period, the average number of observations per user decreased over the years, potentially indicating that the newer users joining the platform contribute fewer observations than existing users. Throughout the years, user activity for CNC iNaturalist also changed in a spatial dimension in Los Angeles, San Francisco and London, starting from an activity focus being centered around the city centre and then spreading out towards the periphery.

Overall, online platforms including citizen science platforms, tend to follow the so-called Nielsen 90-9-1 rule for ratio of activity types (Bégin, Devillers and Roche 2018; Gasparini et al. 2020). The 90-9-1 rule states that 90% of users are 'lurkers' who almost never contribute to generate content, 9% of users provide only minor contributions, and 1% of users, referred to as superusers, account for almost all the contributions. Similarly, the Pareto rule (also called 80/20 rule) is present in some open source communities, which observes that a majority of contributions (80%) tend to be produced by a small subset of the developers (20%), known as the core team (Bégin, Devillers and Roche 2018). This is the case in iNaturalist as well, where 80% of the observations - are made by the top 20% of users (**Figure 3 b).** A scale-free distribution of users' activity on the platform is often observed in common social network platforms such as facebook and others (Aristeidou, Scanlon and Sharples 2015; Barabási 2003). Similar pattern of participation inequality in which a (very) small group of participants contribute most of the data has been found across many citizen science projects (Haklay 2016).

The analysis of the users activity trends can potentially help citizen science platforms such as iNaturalist to understand their communities and introduce relevant systems for the sustainable growth of their communities. We propose a new type of user classification (low activity, identifiers, observers, high activity), which could be applied to a wider range of contexts of citizen science and other platforms with two different potential contribution types. Such classification can help to characterise quantitative changes in the user participation in the platform and as a result, can prevent users' attrition and understand the citizen science project evolution. Another relevant question regarding user's behaviour to analyze in the future is the prediction of user's attrition using special survival analyses or Kaplan-Meier analysis (Bégin, Devillers and Roche 2018) for identification of phases in contributors' life cycle, as well as the characterisation of stages of contribution typical of participation in CS projects (Fischer, Cho and Storksdieck 2021).

We used data science tools to explore the patterns formed by iNaturalist users that participated in CNC across three cities. We observe the intricate growth of the iNaturalist community over time and also spatially. There are many sides to the growth of a community: the community grows, increasing the territory it covers spatially, while decreasing the number of connections within the community, as well as changing the core participants, who contribute to the platform. The analysis of a social network uncovered hidden patterns of users activity - the community of iNaturalist sustains itself not just through new observations in new areas of the city, but importantly through creating links between users. Network science can offer many more advanced tools and methods for social network analysis, promising many more insights. The iNaturalist network structure could be further examined by investigating structural properties such as spatial communities per location or topics, as well as a temporal dimension of network formation. Overall, the computational methods and analysis we propose in the paper enabled the new insights into the dynamic interaction processes and reinforcement activities between users in citizen science iNaturalist platform.

**Acknowledgements**

We thank Frie Preu and the CorrelAid team, without whom this project would not have been possible. We also thank the iNaturalist community, who graciously took time to explain some of their data to us and discussed the results of this work with us.

**Competing Interests Statement**

The authors have no competing interest to declare.

**Authors' Contributions Statement**

FS, VV and LT made the analyses regarding social analysis and users' attrition, KK worked on the general statistics, and FL and AM on the spatial analyses. All authors contributed to a first draft of the paper.  BGT, MH and LT organised and settled the project, and participated in framing and writing the paper. KK and LT made the final formatting.

**Funding Information**

The project was done as well as coordinated on a voluntary basis. Authors acknowledge financial support for publication from the Center for Research and Interdisciplinarity (CRI) and from University College of London (UCL).

**Figures and Captions**

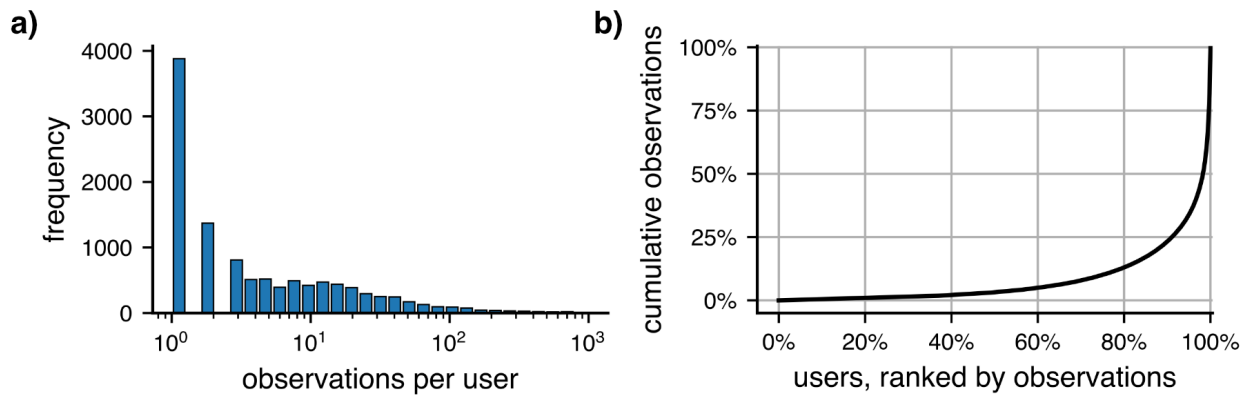

**Figure 1a** The distributions of the count of observations per user over all CNCs. Users who participated in several events were counted as new participants in each CNC. The horizontal axis denotes the number of observations, the vertical axis denotes the number of people who made such a number of observations. **b** The cumulative number of observations across fractions of the user base. Users were sorted by their number of observations in ascending order. Most observations are contributed by few high-contributing users.

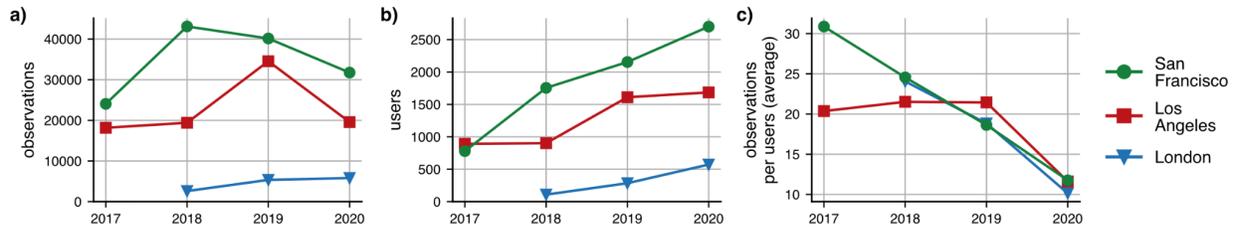

**Figure 2** Comparison of user contributions across cities and years. **a** The number of observations, **b** the number of users, and **c** the average number of observations per user over the years across different cities: San Francisco (green, circle), Los Angeles (red, square) and London (blue, triangle).

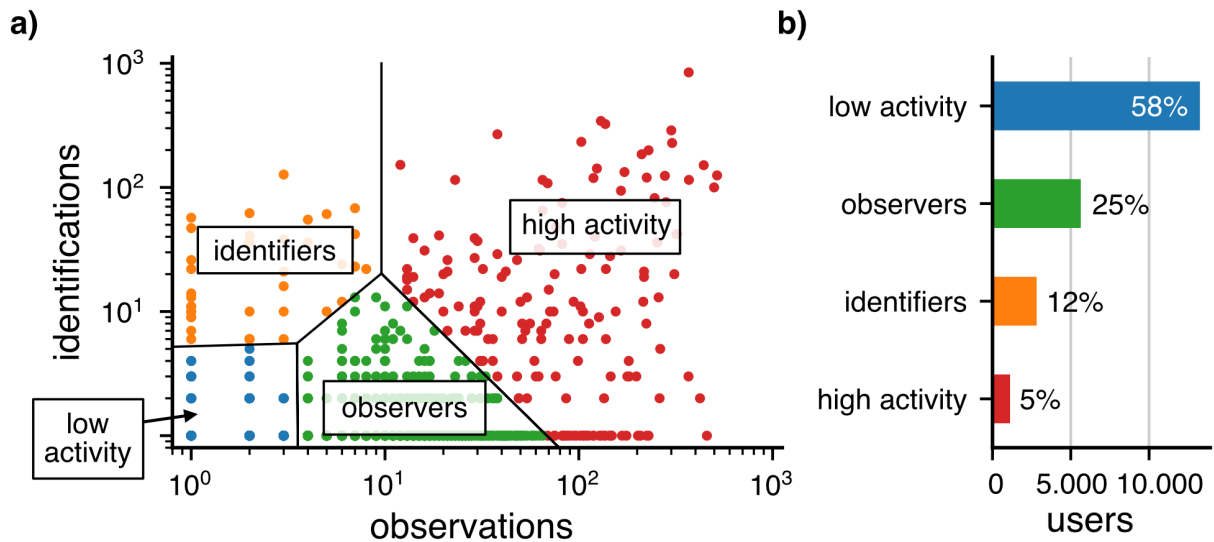

**Figure 3** Users classification. **a** We group users into four clusters, depending on their behavior during the CNC: low activity users (blue), who only added few observations or identifications; observers, who mainly submitted observations (green); identifiers, who mainly contributed identifications (orange), and high activity users (red), who both identified and observed in high quantity. Shown here is the classification of users for the

San Francisco 2020 CNC. **b** According to this classification most users are low activity users (58%), while only few users (5%) have many interactions.

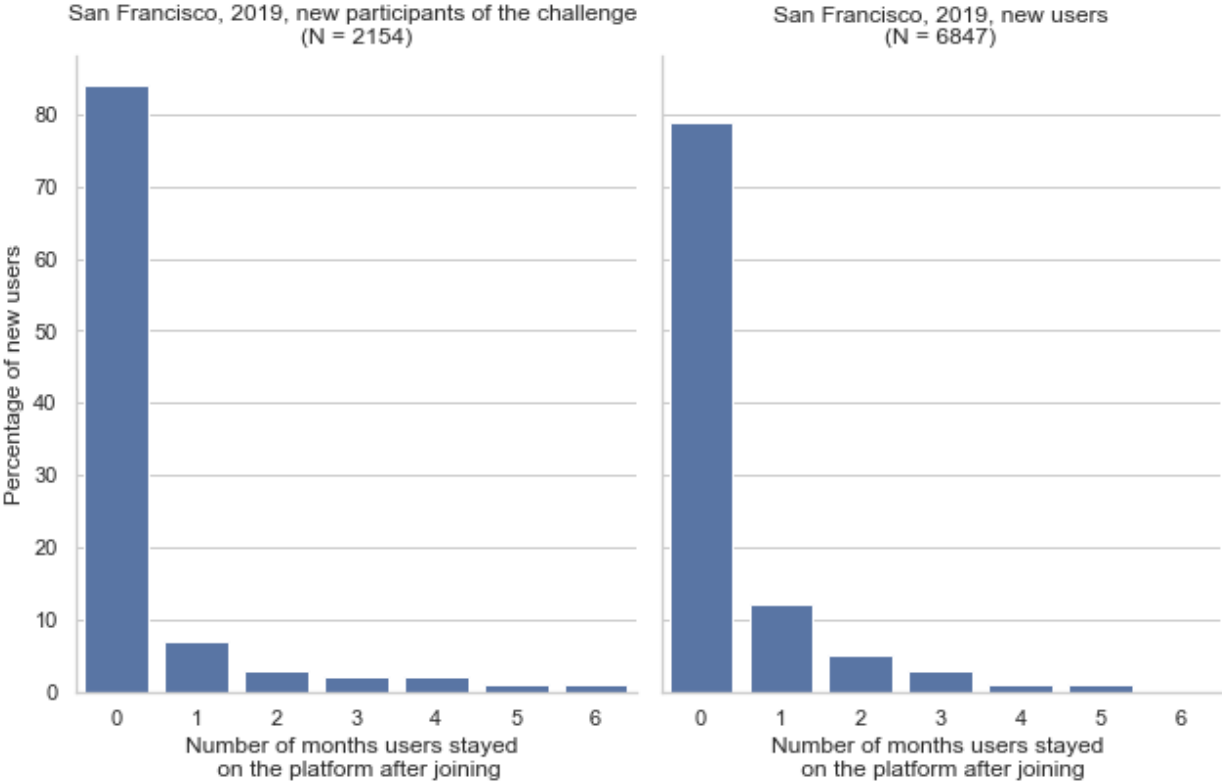

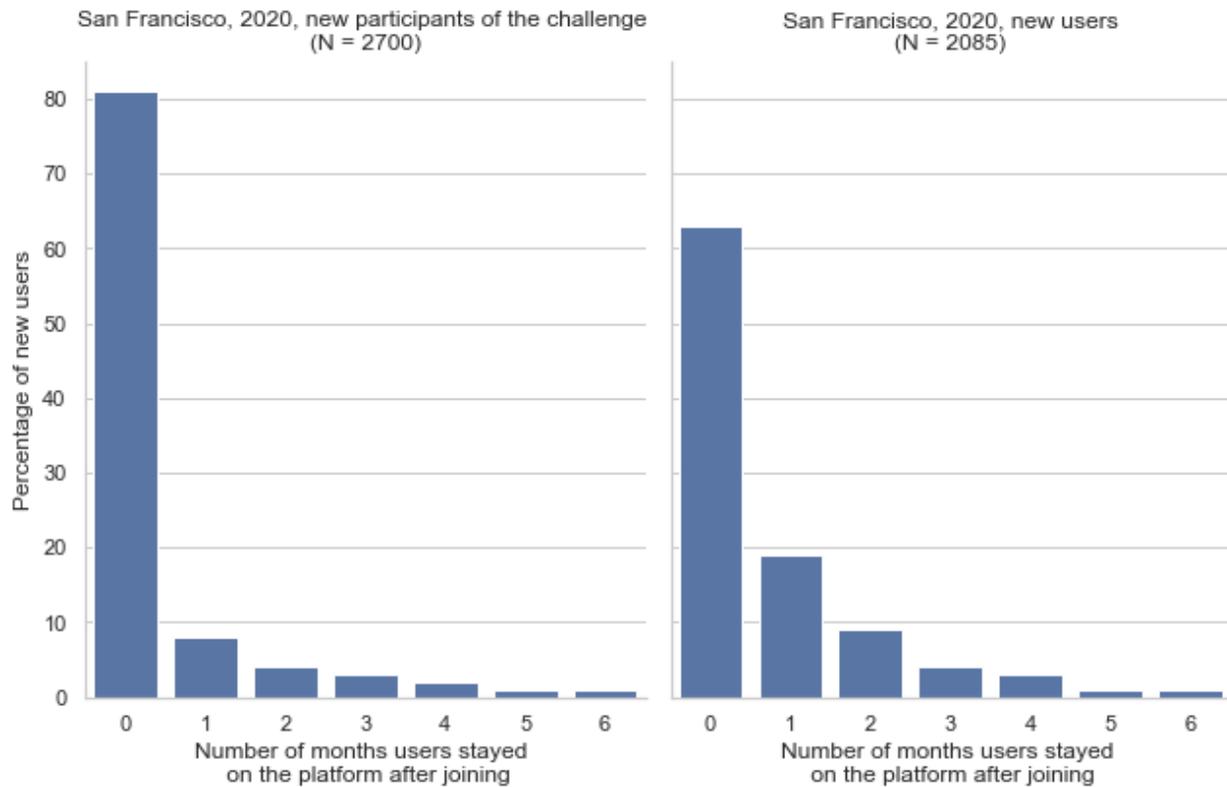

**Figure 4** the attrition of users in San Francisco in 2019 for CNC participants (left), for regular platform users (right). **b** The attrition of users in San Francisco in 2020 for CNC participants (left), for regular platform users (right).

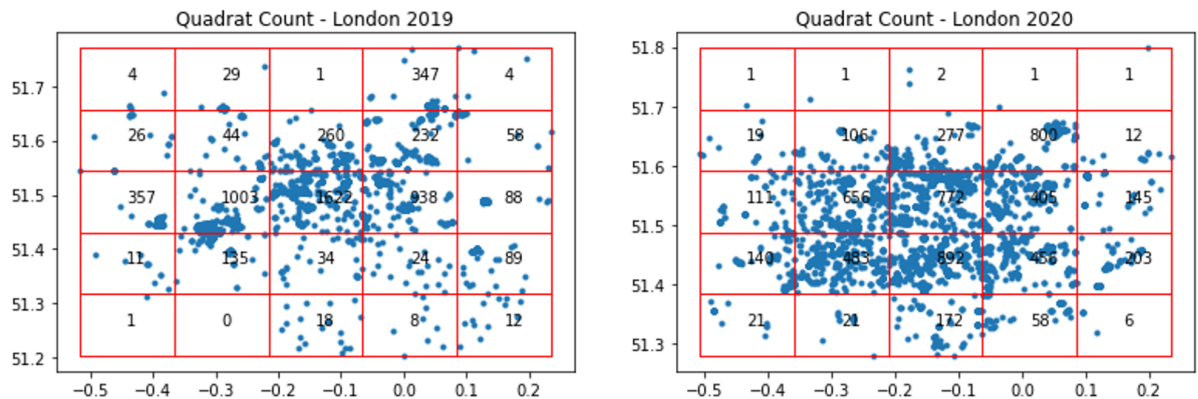

**Figure 5** The distribution of observations in London made in 2019 (**left panel**), depicted as a quadrant count and in 2020 (**right panel**).

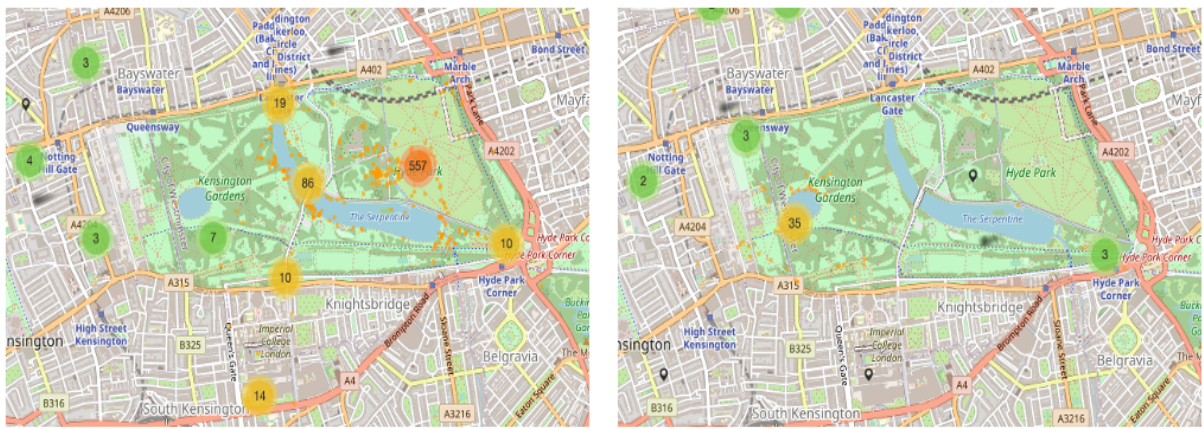

**Figure 6** The distribution of observations in London Hyde Park made in 2019 (**a**) and in 2020 (**b**). The number of observations in the park clearly decreased. For more interactive plots, see [Correlaid inaturalist website].

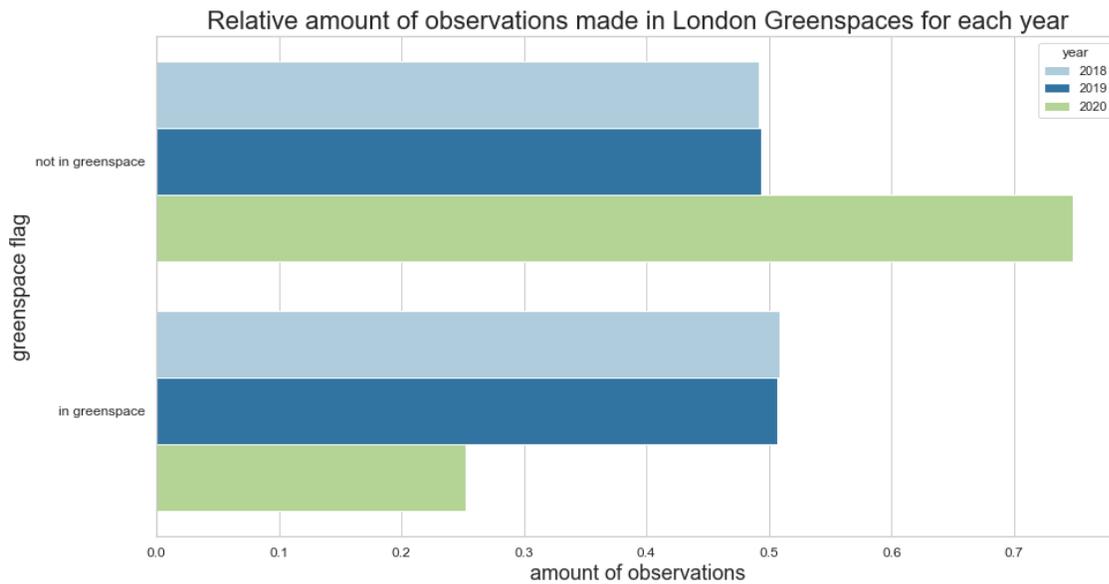

**Figure 7** Relative amount of observations made in London greenspaces for each year.

The number of observations in greenspaces decreased substantially in 2020.

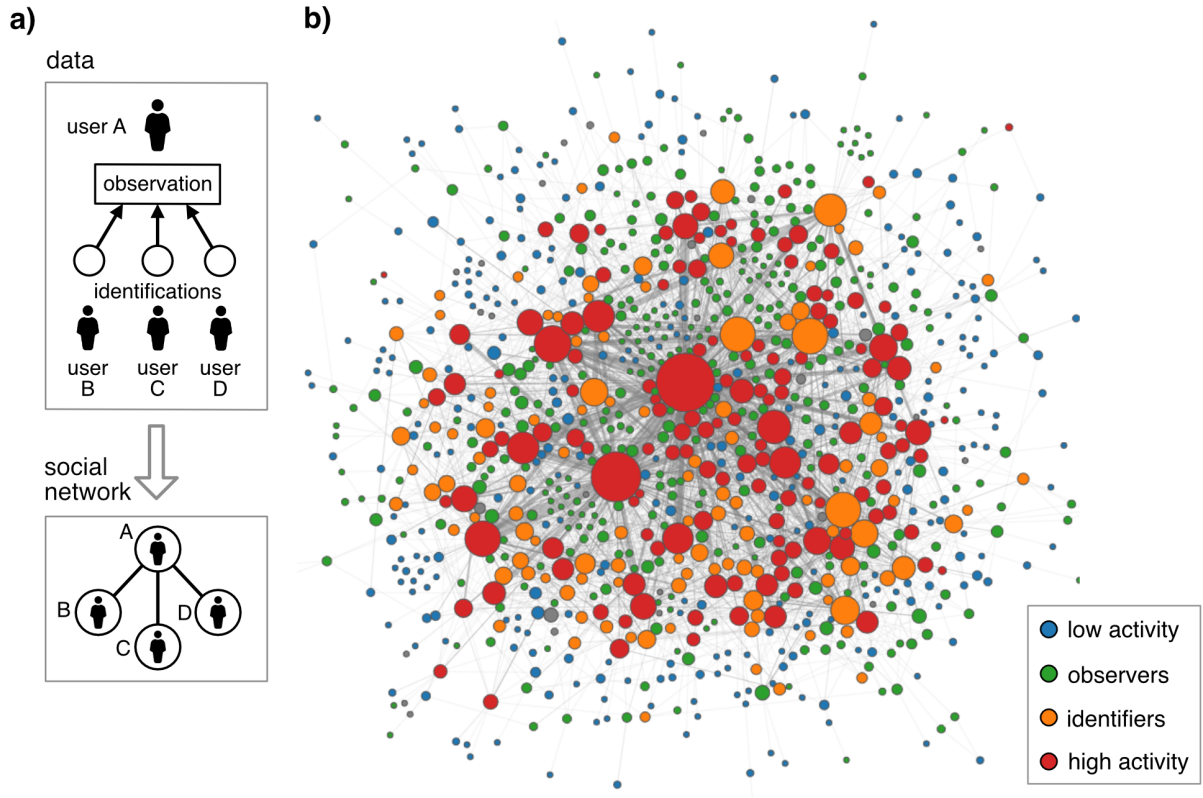

**Figure 8** Illustration of the social interaction network. **a.** A link between two users is created if a user identifies an observation of another user. Multiple interactions increase the link weight. **b** The social interaction network for the London 2020 CNC. Colors

indicate the users' cluster. Links are sized according to their weight, nodes according to the sum of all in- and outgoing weight. High activity users (red) and identifiers (orange) are most prominent and central in the network, while low activity users (blue) and observers (green) are located towards the periphery. The graph was visualized using the Netwulf Python package [Netwulf].

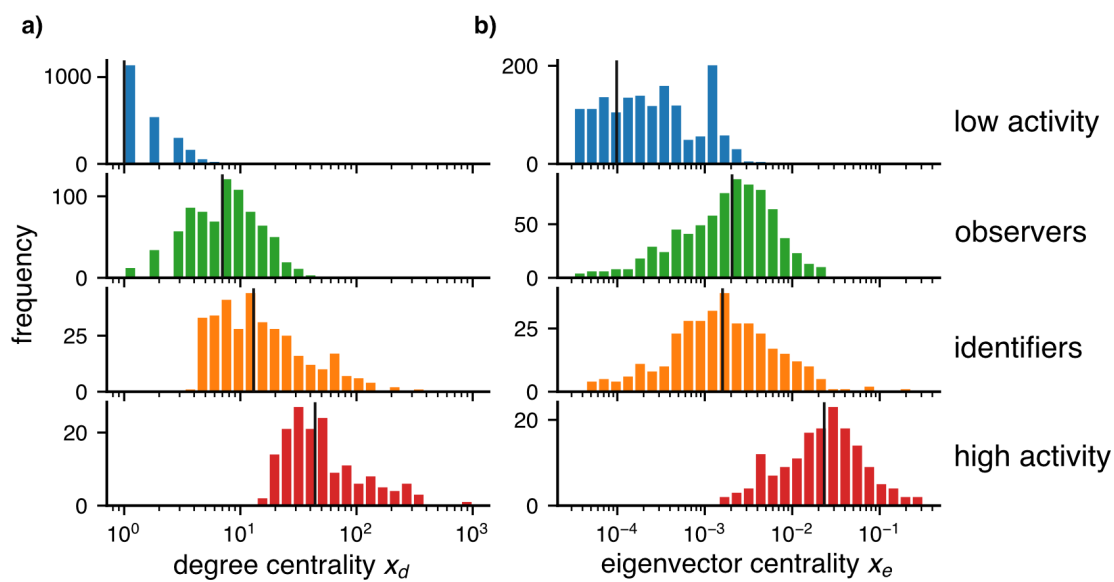

**Figure 9** Histograms of the **a** degree centrality and **b** eigenvector centrality of users in the social interaction network for the San Francisco 2020 CNC, where users are grouped according to their cluster. The black lines indicate the median.